\documentclass[times,twocolumn,final,longtitle]{elsarticle}

\usepackage{medima}
\usepackage{framed,multirow}
\usepackage{subcaption}
\usepackage[dvipsnames]{xcolor}
\usepackage[unicode=true]{hyperref}

\DeclareUnicodeCharacter{9225}{}

\usepackage{amssymb}
\usepackage{latexsym}
\usepackage{algorithmic}
\usepackage{amsfonts}
\usepackage{amsmath}
\usepackage{bbding}

\usepackage{url}
\usepackage{booktabs}
\usepackage{makecell}
\usepackage{natbib}
\journal{Medical Image Analysis}
\begin{document}

\verso{Meiyue Song \textit{et~al.}}

\begin{frontmatter}

\title{PneumoLLM: Harnessing the Power of Large Language Model for Pneumoconiosis Diagnosis}%

\author[1,2]{{Meiyue \snm{Song}$\dagger$}}
\author[3]{{Jiarui \snm{Wang}$\dagger$}}
\author[4]{{Zhihua \snm{Yu}$\dagger$}}
\author[5]{{Jiaxin \snm{Wang}$\dagger$}}
\author[6]{{Le \snm{Yang}$\dagger$}}
\author[3]{{Yuting \snm{Lu}}}
\author[7]{{Baicun \snm{Li}}}
\author[8,9]{{Xue \snm{Wang}}}
\author[3]{{Xiaoxu \snm{Wang}}}
\author[10]{{Qinghua \snm{Huang}}}
\author[11,12]{{Zhijun \snm{Li}}}
\author[13,14,15]{{Nikolaos \snm{I.Kanellakis}}}
\author[1,16,17]{{Jiangfeng \snm{Liu}}\corref{cor1}}
\ead{ljf@pumc.edu.cn}
\author[1,2]{{Jing \snm{Wang}}\corref{cor1}}
\ead{wangjing@ibms.pumc.edu.cn} 
\author[3]{{Binglu \snm{Wang}}\corref{cor1}}
\ead{wbl921129@gmail.com}
\author[1,16,17]{{Juntao \snm{Yang}}}

\cortext[cor1]{Corresponding author.\\$\dagger$Authors contributed equally.}
\address[1]{Institute of Basic Medical Sciences Chinese Academy of Medical Sciences, School of Basic Medicine Peking Union Medical College, Beijing, 100005, China}
\address[2]{State Key Laboratory of Respiratory Health and Multimorbidity, Beijing, 100005, China}
\address[3]{School of Automation, Northwestern Polytechnical University, Shaanxi, Xi’an 710072, China}
\address[4]{Jinneng Holding Coal Industry Group Co.Ltd Occupational Disease Precaution Clinic, Shanxi, 037001, China}
\address[5]{School of Medicine, Tsinghua University, Beijing, 100084, China}
\address[6]{School of Electronics and Control Engineering, Chang’an University, Shaanxi, Xi’an 710064, China}
\address[7]{Center of Respiratory Medicine, China-Japan Friendship Hospital, National Center for Respiratory Medicine, Institute of Respiratory Medicine, Chinese Academy of Medical Sciences, National Clinical Research Center for Respiratory Diseases, Beijing, 100020, China}
\address[8]{Department of Respiratory, the Second Affiliated Hospital of Harbin Medical University, Harbin, Heilongjiang, 150086, China}
\address[9]{Internal Medicine, Harbin Medical University, Harbin, Heilongjiang, 150081, China}
\address[10]{School of Artificial Intelligence, OPtics and ElectroNics (iOPEN), Northwestern Polytechnical University, Xi'an 710072, China}
\address[11]{Translational Research Center, Shanghai YangZhi Rehabilitation Hospital (Shanghai Sunshine Rehabilitation Center), Shanghai 201619, China}
\address[12]{School of Mechanical Engineering, Tongji University, Shanghai 201804, China}
\address[13]{Laboratory of Pleural and Lung Cancer Translational Research, CAMS Oxford Institute, Nuffield Department of Medicine, University of Oxford, Oxford, UK}
\address[14]{Oxford Centre for Respiratory Medicine, Churchill Hospital, Oxford University Hospitals NHS Foundation Trust, Oxford, UK}
\address[15]{National Institute for Health Research Oxford Biomedical Research Centre, University of Oxford, Oxford, UK}
\address[16]{Plastic Surgery Hospital, Chinese Academy of Medical Sciences and Peking Union Medical College, Beijing, 100144, China}
\address[17]{State Key Laboratory of Common Mechanism Research for Major Diseases, Beijing, 100005, China}

\received{XXX}
\finalform{XXX}
\accepted{XXX}
\availableonline{XXX}
\communicated{XXX}

\begin{abstract}
The conventional pretraining-and-finetuning paradigm, while effective for common diseases with ample data, faces challenges in diagnosing data-scarce occupational diseases like pneumoconiosis. Recently, large language models (LLMs) have exhibits unprecedented ability when conducting multiple tasks in dialogue, bringing opportunities to diagnosis. A common strategy might involve using adapter layers for vision-language alignment and diagnosis in a dialogic manner. Yet, this approach often requires optimization of extensive learnable parameters in the text branch and the dialogue head, potentially diminishing the LLMs' efficacy, especially with limited training data. In our work, we innovate by eliminating the text branch and substituting the dialogue head with a classification head. This approach presents a more effective method for harnessing LLMs in diagnosis with fewer learnable parameters. Furthermore, to balance the retention of detailed image information with progression towards accurate diagnosis, we introduce the contextual multi-token engine. This engine is specialized in adaptively generating diagnostic tokens. Additionally, we propose the information emitter module, which unidirectionally emits information from image tokens to diagnosis tokens. Comprehensive experiments validate the superiority of our methods.
\end{abstract}

\begin{keyword}
\KWD Large language model\sep Medical image diagnosis\sep Foundational model
\end{keyword}

\end{frontmatter}


\section{Introduction}
In the computer-aided diagnosis community, the processing and analysis prowess applied to medical data is pivotal. It facilitates the diagnosis of potential diseases and the prediction of future clinical outcomes. With the rapid evolution of deep learning theories, researchers have designed sophisticated network architectures \citep{he2016deep, dosovitskiy2020image} and have curated extensive, high-quality datasets \citep{deng2009imagenet, wang2017chestx} to pretrain these powerful networks. Pretraining strategies endow networks with valuable knowledge by optimizing weight distribution, which, in turn, equips researchers to further refine the model with labeled data targeted at diagnosing specific diseases. When the data is abundant and accurately labeled, this classical paradigm typically achieves commendable results, particularly with common ailments. An example is EchoNet-Dynamic \citep{ouyang2020video}, which has surpassed medical specialists in cardiac function assessment.
\begin{figure*}[htbp]
\centering
\includegraphics[width=0.8\linewidth]{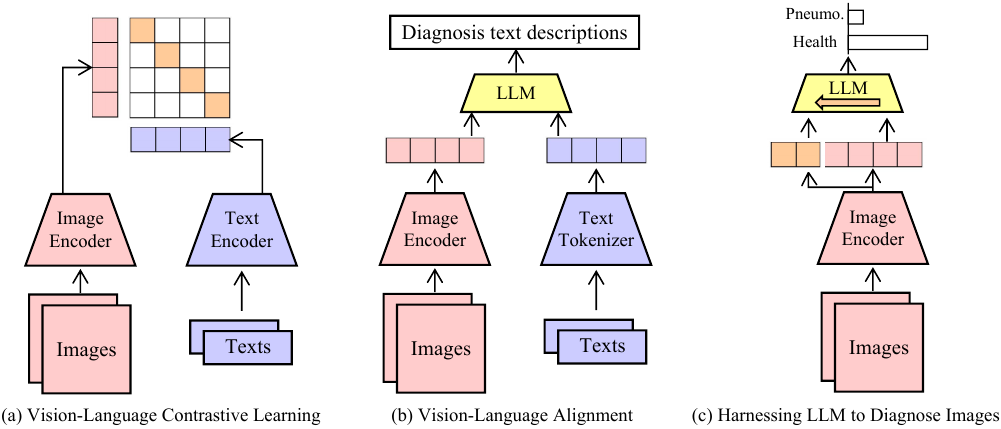}
\caption{Representative pipelines to elicit knowledge from large models. (a) Traditional works conduct vision-language contrastive learning to align multimodal representations. (b) To utilize large language models, existing works transform images into visual tokens, and send visual tokens to LLM to generate text descriptions. (c) Our work harnesses LLM to diagnose medical images by proper designs, forming a simple and effective pipeline.}
\label{figMotivation}
\end{figure*}

However, the landscape shifts when we delve into occupational diseases such as pneumoconiosis \citep{li2023potential, dong2022use}. Individuals subjected to prolonged exposure in dust-laden environments without personal protective equipments are susceptible to pulmonary fibrosis, a precursor to pneumoconiosis \citep{qi2021pneumoconiosis, devnath2022detection}. Areas with increased prevalence of pneumoconiosis often lack economical development, medical resources and infrastructure, and prefessional medical practitioners. Furthermore, there is a noticeable reticence towards disease screening and diagnosis, leading to an acute shortfall in clinical data for these diseases \citep{sun2023expertnet, huang2023association}. This paucity of data renders the conventional pretraining-and-finetuning strategy ineffective.

Addressing the diagnostic challenges posed by data-scarce occupational diseases requires an inventive approach. It involves tapping into the rich knowledge \citep{huang2023novel} of foundational models and controlling the amount of learnable parameters to streamline the learning process. The advent of large language models (LLMs) \citep{kenton2019bert, brown2020language} has gained a bounty of knowledge from voluminous pretraining corpora, showcasing an impressive generalization capability for new tasks. The medical image diagnosis community has witnessed the emergence of foundational models \citep{zhang2023challenges, zhang2024data}, with significant strides made in pathology image analysis \citep{zhang2023text, huang2023visual} and medical image segmentation \citep{cheng2023sam, lei2023one, wang2023sam}.

In the wake of such LLMs breakthroughs, concerted efforts continue to leverage knowledge from large-scale models to enhance image processing tasks, as depicted in Fig. \ref{figMotivation}. For instance, CLIP \citep{radford2021learning} embarks on vision-language contrastive learning to carefully align visual and language representations, as showcased in Fig. \ref{figMotivation}(a). The multimodal community, in turn, benefits from the integration of LLMs by interpreting visual tokens as a specialized form of language and devising adapters \citep{li2023blip2, zhang2023LLaMA} to convert visual inputs into comprehensible representations, as shown in Fig. \ref{figMotivation}(b). These works often employ advanced techniques, such as instruction tuning \citep{stiennon2020learning, liu2024visual}, to yield fluent and varied narrative outputs. Meanwhile, some works begin to explore the interactive tool for interpreting the inner attention mechanisms of large vision-language models \citep{stan2024lvlm}. The medical image diagnosis sector has also made notable advances by developing vision-language models \citep{wen2023improving, yi2023towards} or by constructing medical foundational models from scratch \citep{moor2023foundation}. \cite{xu2023learning} propose a unified transformer model specifically designed for multi-modal clinical tasks by incorporating customized instruction tuning. \cite{li2024llava} introduce conversational generative AI into the biomedicine domain and can follow open-ended instruction to assist with inquiries about a biomedical image. 

Nonetheless, the application of these existing LLM-based pipelines to diagnose data-scarce occupational diseases poses several challenges. Firstly, the dependency on ample paired image-text data intensifies the complexity of data gathering, particularly when factoring in the constraints of patient privacy. Besides, processing text inputs through separate branches escalates computational demands substantially. Although the textual outputs are versatile, they may be unnecessarily complex for tasks that require simple binary outcomes, such as affirming the presence or absence of a specific disease.

Our approach diverges from existing methodologies by \textit{eliminating the textual processing branch and directly harnessing LLMs to process images for the diagnosis of pneumoconiosis}, as shown in Fig. \ref{figMotivation}(c). We hypothesize that LLMs, after extensive corpus learning, are adept at selecting salient visual tokens and filtering out irrelevant ones, thereby benefiting the medical image diagnostic process. Fig. \ref{figFramework} presents the proposed PneumoLLM framework. We revise the language prediction head into a classification head, transitioning from dialogue-based outputs to succinct disease classification. After freezing parameters of both the vision encoder and the LLM, we integrate the adapter module and manage the learnable parameters effectively. We ascertain that eliciting diagnostic knowledge from LLMs hinges on balancing the preservation of comprehensive image details with the progression towards specific diagnostic task. To navigate this balance, we introduce the contextual multi-token engine that generates diagnostic tokens conditioned on image tokens. This ensures that the source image tokens retain all the pertinent image details. Subsequently, the information emitter module is engineered to unidirectionally emit information from source tokens to diagnosis tokens, thus steering the learning trajectory towards accurate diagnosis. All code is available at https://github.com/CodeMonsterPHD/PneumoLLM.

In brief, this work contributes to the field by:
\begin{itemize}
    \item Charting new paradigm in applying LLMs to medical image analysis, especially for data-scarce occupational diseases, thereby simplifying the diagnostic process while yielding promising results.
    \item Designing the novel contextual multi-token engine and information emitter module to meticulously draw out knowledge from LLMs, achieving a harmonious balance between preserving image representations and harnessing LLMs diagnostic intelligence.
    \item Demonstrating the superiority of our method in diagnosing pneumoconiosis through extensive experiments and validating the effectiveness of each designed module.
\end{itemize}

\section{Related Work}

\subsection{Disease diagnosis based on X-ray}
Recent advancements in deep learning have shown significant promise in the field of medical image diagnosis. \cite{luo2022word} establishd a large-scale whole abdominal organ Dataset (WORD) for deep learning-based algorithm research and clinical application development. \cite{luo2022scpm} developed a 3D sphere representation-based center-points matching detection network specifically for detecting pulmonary nodules in CT images. \cite{wu2023pattern} proposed a pattern-aware transformer to achieve hierarchical pattern matching and propagation among sequential medical images. \cite{ali2020additive} introduced a novel additive angular metric for few-shot classification of diverse endoscopy samples within a prototypical network framework. \cite{kang2022thyroid} proposed a method for intra- and inter-task consistent learning, enhancing model predictions across various related tasks and addressing inconsistencies inherent in such tasks. 

In the realm of anomaly detection, \cite{li2023self} pioneered an unsupervised framework, SSL-AnoVAE, which leverages self-supervised learning \citep{huang2023stu, wang2023mis} to provide fine-grained semantic analysis for anomalies in retinal images. \cite{huang2022transformer} introduced a transformer-based approach for classifying pneumoconiosis in 3D CT images, effectively combining intra-slice and inter-slice interaction information. \cite{chen2023dynamyic} proposed a dynamic feature splicing strategy for few-shot diagnosis of rare diseases (e.g., hernia), employing similarity channel replacement at both low and high feature levels. \cite{xing2023gradient} presented a two-stage diagnostic framework involving multi-modal learning and cross-modal distillation, addressing challenges of limited dataset size and structural variations. \cite{xu2023clinically} combined deep learning and machine learning methods for segmentation and feature extraction, mimicking the workflow of experienced radiologists. \cite{chen2023orthogonal} introduced OLFG, an orthogonal latent space learning approach with feature weighting and graph learning for multimodal Alzheimer's Disease diagnosis. \cite{ma2023multi} developed MGCA-RAFFNet, a multi-graph cross-attention-based network for brain disorder diagnosis, utilizing multi-template analysis. \cite{fan2023one} extended conventional siamese networks for low-shot learning, introducing a semi-supervised strategy that utilizes unlabeled data to enhance accuracy.

As for disease diagnosis based on X-ray images, \cite{wang2020covid} proposed COVID-Net, a ResNext50 network pre-trained on ImageNet and employing a lightweight PEPX design pattern. \cite{zheng2019improved} restructured GoogLeNet using convolutional kernel decomposition. \cite{wang2020potential} utilized GoogleNet (Inception-v3) to detect pneumoconiosis. \cite{devnath2021automated} employed two Convolutional Neural Network (CNN) models for feature extraction in pneumoconiosis CR images. \cite{gao2021covid} developed a vision transformer based on attention models and DenseNet for COVID-19 classification from 2D slices of 3D CT images. \cite{heidarian2021cae} proposed a CAE-Transformer framework for efficient classification of lung adenocarcinoma tumours using whole 3D CT images. The methods for diagnosing pneumoconiosis are summarized in Tab. \ref{relatedwork}. However, the reliance on data-driven deep learning methods necessitates ample training data, presenting challenges for occupational diseases like pneumoconiosis.

\begin{table*}[htbp]
  \centering
  \small
  \caption{Existing diagnosis methods for pneumoconiosis.}
    \begin{tabular}{c|c|c}
    \toprule
    Method & Advantages & Disadvantages \\
    \midrule
    \midrule
    \cite{huang2022transformer}& \makecell{Captured intra-slice dependencies and \\ inter-slice information exchange.} & \makecell{Large number of model parameters.} \\
    \midrule
    \cite{zheng2019improved}& \makecell{Modeled features at different scales.} &\makecell{Complex network structure that is \\ prone to overfitting on small datasets} \\
    \midrule
    \cite{wang2020potential} & \makecell{Improved the network's ability to learn \\ features at different scales and positions.} & \makecell{High training complexity and complex \\ parameter configuration.} \\
    \midrule
    \cite{devnath2021automated} & \makecell{Used a multi-layer feature aggregation method to \\ address the dust lung disease detection on a small dataset.} & {Sensitive to hyperparameter selection.} \\
    \bottomrule
    \end{tabular}%
    \label{relatedwork}%
\end{table*}%

\subsection{Foundational models and applications to diagnosis}

The emergence of foundational models in the natural language domain, exemplified by the pre-training of Transformer \citep{vaswani2017attention} and BERT \citep{kenton2019bert}, has demonstrated remarkable generalization abilities. Researchers have developed various parameter-efficient fine-tuning strategies, such as prefix tuning \citep{lester2021power} and adapter methods \citep{houlsby2019parameter}, to leverage the potential of these models, often achieving competitive or superior performance in downstream tasks. The advent of CLIP, through its contrastive pre-training approach, has established a robust vision-language foundational model \citep{radford2021learning}. This development has significantly advanced zero-shot and few-shot learning tasks, facilitated by innovative prompt tuning strategies \citep{zhou2022learning} and adapter techniques \citep{hu2021lora}.

Building upon the aforementioned progress in vision-language models, BLIP-2 \citep{li2023blip2} integrates pre-existing vision and language models, freezing the original parameters while learning an additional transformation network, thereby generating strong vision-language representations. In the vision domain, recent advancements in foundational models have focused on providing general representations for a variety of downstream tasks \citep{dinov2} and enhancing performance in open-world environments, including segmenting any object \citep{kirillov2023segment} and recognizing diverse entities \citep{zhang2023recognize}. In the language domain, development efforts have led to the creation of large-scale foundational models, such as PaLM \citep{chowdhery2022palm} and ChatGPT \citep{chatgpt}, which, being non-open-sourced, are accessible only through APIs. Conversely, other efforts have produced open-sourced models like LLaMA \citep{touvron2023LLaMA}, opening new avenues for research.

The medical image diagnosis domain has also benefitted considerably from foundational models \citep{gao2023training}. For example, Med-PaLM \citep{singhal2023large} and DoctorGLM \citep{xiong2023doctorglm} infuse extensive medical knowledge into general foundational models. Similarly, MedCLIP \citep{wang2022medclip} and CXR-CLIP \citep{you2023cxr} utilize X-ray images to pretrain foundational models specialized for disease diagnosis. Subsequent research has focused on exploring and harnessing the rich knowledge embedded in these models, developing disease-specific adaptations through methods like prompt-tuning \citep{zhang2023text}, adaptation \citep{wang2023real}, and continual learning \citep{yi2023towards}. Additionally, efforts have been made to develop multimodal foundational models, such as PLIP \citep{huang2023visual} and RadFM \citep{wu2023towards}, targeting a wide array of diagnostic tasks in a unified manner.

Despite the promising potential of foundational models, they typically require a substantial volume of paired image-text training data and often generate predictions in a dialogue format. In the context of pneumoconiosis diagnosis, the available data is limited to hundreds of images, and the annotations are classification labels rather than dialogue sentences. Therefore, this work represents an early exploration into harnessing the rich knowledge within foundational language models for direct application to image diagnosis tasks.

\section{Method}
\subsection{Overview}
The efficacy of computer-aided diagnosis systems is crucial in processing and analyzing medical data. However, these systems often face a significant shortfall in clinical data availability. Leveraging the rich knowledge reservoirs of foundational models is a promising strategy to address this data scarcity. Yet, the conventional pretraining-and-finetuning approach may compromise the representation capabilities of LLMs, due to substantial changes in their parameter spaces, leading to increased training time and memory overhead \citep{touvron2023LLaMA, touvron2023LLaMA2, openai2023gpt4}.

\begin{figure*}[thbp]
\centering
\includegraphics[width=1\linewidth]{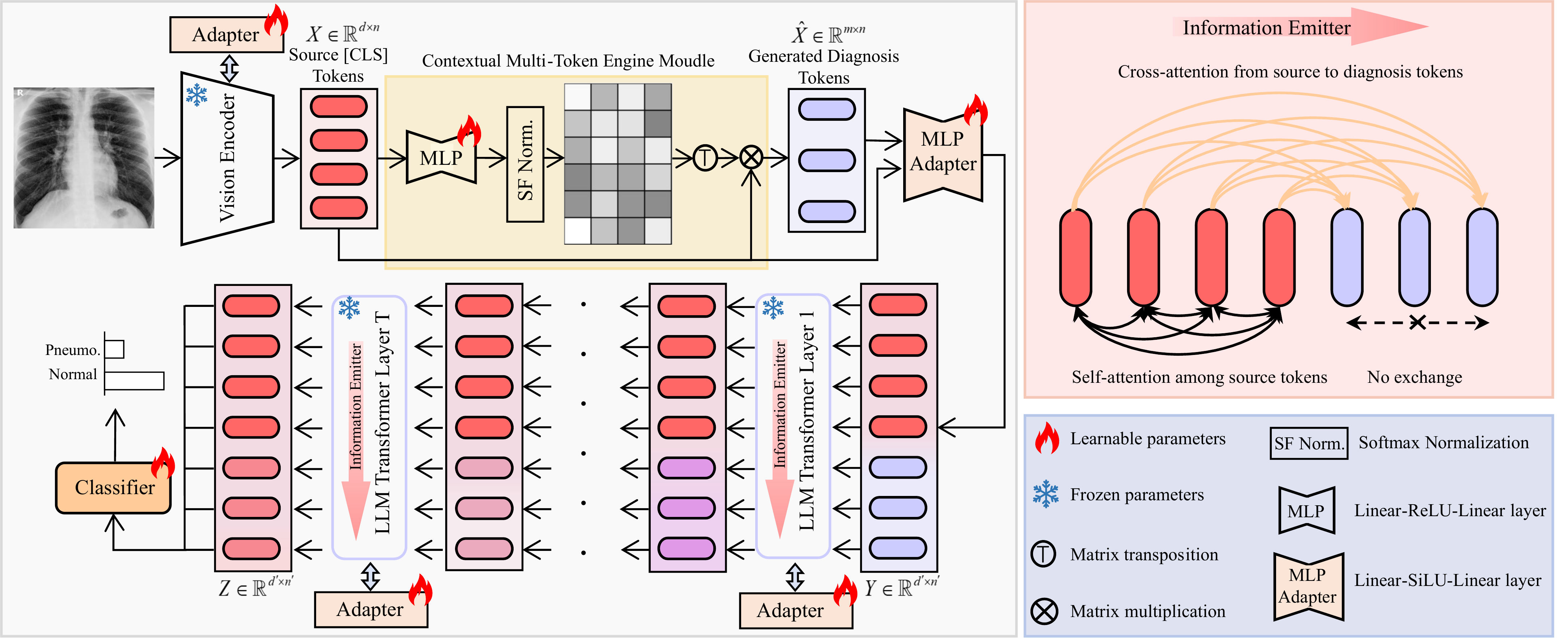}
\caption{Diagram of the proposed PneumoLLM. The vision encoder processes chest radiography and extracts source tokens. The contextual multi-token engine generates multiple diagnosis tokens conditioned on source tokens. To elicit in-depth knowledge from the LLM, we design the information emitter module within the LLM Transformer layers, enabling unidirectional information flow from source tokens to diagnosis tokens, preserving complete radiographic source details and aggregating critical diagnostic information.}
\label{figFramework}
\vspace{-0.3cm}
\end{figure*}

To mitigate these challenges, we introduce PneumoLLM, an innovative LLM-based framework tailored for diagnosing pneumoconiosis using chest radiographs. PneumoLLM begins by processing chest radiographs (Fig. \ref{figFramework}) through a vision encoder to extract informative source tokens, which are subsequently input into the LLMs to derive the final classification results. To ensure effective integration between the vision encoder and the LLMs, as well as to enhance the adaptability of the LLM to specific diagnosis task while preserving its original structure, we propose the integration of additional diagnosis tokens. To achieve this, we introduce two modules: contextual multi-token engine and information emitter modules. The former is responsible for generating the additional contextual diagnosis tokens, while the latter emits information from source tokens to additional contextual diagnosis tokens, preserving complete radiography source details and consolidating valuable diagnostic information. Additionally, to avoid disrupting the LLM's robust representation, we introduce adapter layers in both the vision encoder and the LLM model.

In detail, a chest radiography image, denoted as \(I_{\rm{img}}\), is processed through a vision encoder \(f_{\rm{vis}}\). We utilize the image encoder from the pretrained CLIP-ViT \citep{radford2021learning} to capitalize on its intrinsic alignment with language data, thereby facilitating the comprehension by subsequent LLMs. The visual features, represented as \(X \in \mathbb{R}^{d \times n}\), are extracted using \texttt{[CLS]} tokens from every fourth layer of the ViT, where $d$ is the number of extracted \texttt{[CLS]} tokens and $n$ is the feature dimension of each token. Subsequently, \(X\) is processed by our contextual multi-token engine module, generating additional contextual diagnosis tokens ${\hat{X} \in \mathbb{R}^{m \times n}}$, where $m$ is the number of newly generated tokens. The original \texttt{[CLS]} tokens \(X\) and the new contextual diagnosis tokens \(\hat{X}\) are then concatenated and passed through a simple adapter MLP layer. This layer, comprising a simple two-layer bottleneck structure (Linear-SiLU-Linear) with the hidden layer dimension reduced to 128, transforms the amalgamated visual tokens \(X' = [X,\hat{X}]\) into dimensions compatible with the LLM, resulting in \(Y \in \mathbb{R}^{d' \times n'}\), where $d'=d+m$ is the total number of tokens resulting from the concatenation of \(X\) and \(\hat{X}\), and $n'$ is the feature dimension of each token in LLM.


The processed tokens \(Y\) are input into the pretrained LLM, sans its final classifier layer, to generate the final features \(Z \in \mathbb{R}^{d'\times n'}\). These features are then fed into our disease classification network ${h_\varphi }\left(  \cdot  \right)$, designed for pneumoconiosis diagnosis, yielding the final classification logit scores \(J \in \mathbb{R}^{c}\), where \(c\) denotes the number of classification categories. In line with the previous approaches \citep{luo2023cheap, zhang2023LLaMA}, we integrate the adapter layer $h_{\phi}\left(  \cdot  \right)$ to each multi-head attention module in both the vision encoder and LLM layers. During training, the learnable parameters in the adapter, the contextual multi-token engine module, and the disease classification network undergo training via a cross-entropy loss function, while the rest of the PneumoLLM parameters remain frozen.

\subsection{Contextual multi-token engine}
\label{sec:promote}
In the field of vision-language alignment, some data-efficient approaches like CoOp \citep{zhou2022learning} and CoCoOp \citep{zhou2022conditional} advocate prompt engineering to achieve better performance than fixed hand-crafted prompts, which serves as an inspiration for our design. The promote engineering may also be useful in directly guiding the LLMs to further complement the information of visual tokens beyond mere vision-language alignment. However, CoOp's limitation lies in its uniform prompts for all samples, restricting instance flexibility during inference. CoCoOp extends this by learning a vision-conditional prompt token, moderately enhancing inference adaptability. Yet, this direct addition of fixed and flexible prompts might confuse LLMs and underutilize its potential. Our proposed contextual multi-token engine module aims generate new diagnosis tokens to seamlessly integrate and maximize the utility of information across diverse vision tokens. They will guide LLMs to generate more complementary features, leading to accurate diagnosis.

Specifically, as shown in the left top of Fig. \ref{figFramework}, we employ a two-layer contextual multi-token MLP network (Linear-ReLU-Linear) denoted as ${h_\theta ^1}$, with the hidden layer reducing the input dimension by 12. This network, along with Softmax normalization, is utilized to generate the output contextual attention map $M_c \in \mathbb{R}^{d \times m}$ conditioned on the original \texttt{[CLS]} tokens ${X \in \mathbb{R}^{d \times n}}$, where $m$ is a hyper-parameter representing the number of generated diagnosis tokens. Subsequently, we employ matrix multiplication to compute the output diagnosis tokens ${\hat{X} \in \mathbb{R}^{m \times n}}$. The entire process can be described in Eq. \eqref{eq_contexmap}.
\begin{equation}
\label{eq_contexmap}
M_c = \sigma ({h_\theta ^1 }(X)),\ \hat{X} = M_c^T \cdot X
\end{equation}

\subsection{Information emitter module}
After obtaining the original \texttt{[CLS]} tokens \(X\) and the new contextual diagnosis tokens \(\hat{X}\), we concatenate and process them through a simple adapter MLP layer ${h_\theta ^2}$, aligning the combined tokens \(X' = [X;\hat{X}]\) with LLM-compatible dimensions, resulting in \(Y \in \mathbb{R}^{d' \times n'}\), where $d'=d+m$ is the total number of tokens resulting from the concatenation of \(X\) and \(\hat{X}\), and $n'$ is the feature dimension of each token in LLM. Subsequently, as shown in the right top of Fig. \ref{figFramework}, we develop the information emitter module to preserve the original LLM's information interactions for the source \texttt{[CLS]} tokens, while allowing the newly generated context-diagnostic tokens to extract and repurpose this information, thereby fostering novel insights for the diagnosis task.

Specifically, we improve the attention mechanism in each ViT layer of the LLM, preventing \texttt{[CLS]} token features from being altered by contextual diagnosis token features. We define a self-attention mask $M \in \mathbb{R}^{d' \times d'}$ and configure its values as follows:
\begin{equation}
\label{eq_mask}
{M_{i,j}} = \left\{ \begin{array}{l}
 - \infty ,\;\;{\rm{if}}\;j > d\\
0,\;\;\;\;\;{\rm{otherwise}}
\end{array} \right.
\end{equation}

Subsequently, this mask guides the multihead self-attention process in each ViT layer, as formulated below:
\begin{equation}
\label{eq_qkv}
\begin{array}{l}
Q = W_q Y,\ K = W_k Y,\ V = W_v Y
\\[1mm]
{\rm{Attn}}\left( {Q,K,V} \right) = {{\rm{Softmax}}}\left(  { Q{{ K^T}}/\sqrt {d'} } +M \right) \cdot V
\end{array}
\end{equation}

According to the mask definition in Eq. \eqref{eq_mask}, the attention calculation can be further formulated as:
\begin{equation}
\label{eq_qkvs}
\begin{array}{l}
{\rm{Attn}}_s\left( {Q,K,V} \right) = {{\rm{Softmax}}}\left(  { Q_s{{ K_s^T}}/\sqrt {d'} } \right) \cdot V_s,
\\[1mm]
{\rm{Attn}}_c\left( {Q,K,V} \right) = {{\rm{Softmax}}}\left(  { Q_c{{ K_s^T}}/\sqrt {d'} } \right) \cdot V_s,
\\[1mm]
{\rm{Attn}}\left( {Q,K,V} \right) = \left[ \begin{array}{l}
{\rm{Att}}{{\rm{n}}_s}\left( {Q,K,V} \right)\\
{\rm{Att}}{{\rm{n}}_c}\left( {Q,K,V} \right)
\end{array} \right]
\end{array}
\end{equation}
where ${\rm{Attn}}_s\left( {Q,K,V} \right)$ represents the self-attention fusion resulting from the information in the original \texttt{[CLS]} tokens. $Q_s$, $K_s$, and $V_s$ represent the query, key, and value information extracted from the original \texttt{[CLS]} tokens. ${\rm{Attn}}_c\left( {Q,K,V} \right)$ denotes the cross-attention fusion results, indicating the newly generated context diagnostic tokens learning from the original \texttt{[CLS]} tokens. $Q_c$ represents the diagnostic query tokens, receiving the emitted information from source tokens.

By this design, we can ensure that the newly generated context diagnostic tokens will not affect the self-attention process of the original \texttt{[CLS]} tokens, but absorb emitted information from them and supplement their own information through the setting of mask. Besides, it should be noted that there is no information interaction between the newly generated context diagnostic tokens to ensure the uniqueness and diversity of promotes information.

\subsection{Network training}

During the training phase, we focus on training the adapter layers $h_{\phi}\left(  \cdot  \right)$, the contextual multi-token engine network ${h_\theta ^1}\left(  \cdot  \right)$, the simple adapter MLP layer ${h_\theta ^2}\left(  \cdot  \right)$, and the disease classification network ${h_\varphi }\left(  \cdot  \right)$. Since the diagnosis of pneumoconiosis is a binary single-label task, we directly used binary cross-entropy loss for training:
\begin{equation}
{\cal L_{\rm bce}}\left( R, \hat{R};\phi ,{\theta _1},{\theta _2},\varphi\right) = - R \log(\hat{R}) + (1 - R) \log(1 -\hat{R})
\end{equation}
where ${\cal L_{\rm bce}}$ denotes the binary cross-entropy loss function, $R$ represents the ground-truth pneumoconiosis diagnose labels, $\hat{R}$ represents the predicted results by our complete PneumoLLM framework, and $\phi ,{\theta _1},{\theta _2},\varphi$ are the learnable parameters to be optimized.

Notebly, when extending PneumoLLM to a multi-category classification task, the loss function may need to be replaced with cross-entropy loss. Furthermore, for multi-label classification, where a sample can belong to multiple categories simultaneously or none at all, the loss function might require multiple binary cross-entropy losses or other suitable losses for multi-label problems. After adjusting the loss function and the number of categories in the final classification linear layer, our PneumoLLM can also be directly applied to these tasks.

\section{Experiments}

\subsection{Experimental setups}

\paragraph{Dataset acquisition and splits}
In this study, we utilized the posterior-anterior chest radiograph database from Jinneng Holding Coal Industry Group Co.Ltd Occupational Disease Precaution Clinic, comprising 630 chest radiographs in DICOM format, including 401 pneumoconiosis cases.
\begin{figure}[htbp]
\centering
\includegraphics[width=1\linewidth]{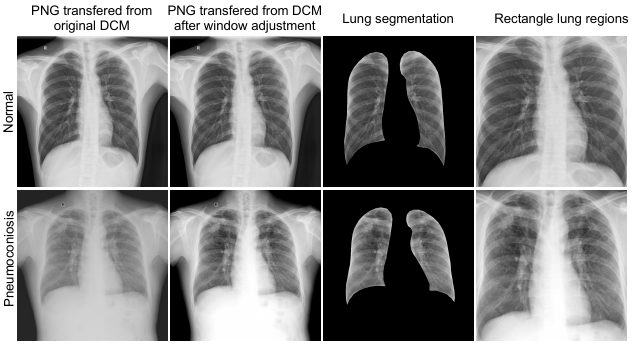}
\caption{The illustration examples of dataset preprocessing: two examples labeled as ``Normal" and ``Pneumoconiosis". The window adjustment opreation use the default window level and width (stored in the DICOM tags) to pre-process the original DICOM files. The segmentation results are obtained using the CheXmask pipeline, as proposed in the paper by \cite{gaggion2023chexmask}. The selection of the rectangular lung regions is based on the largest external rectangle of the segmentation results.}
\label{fig_data_example}
\end{figure}

\begin{table*}[htbp]
  \centering
  \caption{Comparison results with recent prestigious methods on the pneumoconiosis dataset.}
    \begin{tabular}{c|c|cccc|c}
    \toprule
    Method & L-para.(M) & Sens.  (\%) & Spec.  (\%) & Acc. (\%) & AUC (\%) & AVG (\%)\\
    \midrule
    \midrule
    ResNet \citep{he2016deep}& 25.56 & {76.55} & {54.49} & {68.57} & {71.11} & {67.68}\\
    ViT \citep{dosovitskiy2020image}& 88.30&{73.56} & {51.96} & {65.72} & {64.23} & {63.87}\\
    Swin Transformer \citep{liu2021swin} & 28.29 & {74.08} & {52.77} & {66.34} & {67.17} & {65.09}\\
    Conformer \citep{conformer} & 23.25 & {70.81} & {59.88} & {66.82} & {70.10} & {66.90}\\
    ConvNeXt  \citep{convnext}& 88.59 & {69.35} & {62.84} & {66.99} & {67.64} & {66.70}\\
    DINOv2  \citep{dinov2}& 22.06 & {75.56} & {57.23} & {68.89} & {70.40} & {68.02}\\
    VAPFormer \citep{kang2023visual} &1.20  & {63.04}  &  {59.26}   &  {61.15}   & {63.39}  &  {61.71} \\
    \midrule
    {PneumoLLM} & 2.70 & \textbf{80.54} & \textbf{67.66} & \textbf{75.87} & \textbf{78.98} & \textbf{75.76} \\
    \bottomrule
    \end{tabular}%
    \label{tabCmpExp}%
\end{table*}%

\begin{table*}[htbp]
  \centering
  \caption{Comparison results with recent LLM-based methods on the pneumoconiosis dataset.}
    \begin{tabular}{c|c|cccc|c}
    \toprule
    Method & L-para.(M) & Sens.  (\%) & Spec.  (\%) & Acc. (\%) & AUC (\%) & AVG (\%)\\
    \midrule
    \midrule
    Zero-Shot CLIP \citep{radford2021learning} & {----}&\textbf{98.75} & {0} & {62.70} & 57.62 & {54.77} \\
    Linear Probe CLIP \citep{radford2021learning}& 0.0008 &{53.71} & {75.08} & {67.31} & 27.96 & {56.02} \\
    CoOp \citep{zhou2022learning}& 0.003&{75.05} & {51.10} & {66.35} & 67.90  & {65.10} \\
    CoCoOp \citep{zhou2022conditional} &0.078&{74.58} & {60.67} & {69.53} & 71.14 & {68.98} \\
    LaVIN \citep{luo2023cheap} & 3.77&{87.80} & {47.52} & {73.33} & 71.82 & {70.12} \\
    BLIP2 \citep{li2023blip2} &107.13& {69.88} & \textbf{69.47} & {69.70 } & 77.90  & {71.12} \\
    \midrule
    {PneumoLLM} & 2.70&80.54 & {67.66} & \textbf{75.87} & \textbf{78.98} & \textbf{75.76} \\
    \bottomrule
    \end{tabular}%
  \label{tab_LLMcom}%
\end{table*}%

To ensure a balanced ratio of pneumoconiosis and normal samples in training and testing, we conduct five-fold cross-validation for experiments. Subsequently, we merge them into five distinct randomized datasets by patients (Datasets 1-5), and report the average performance of five experiments. Each dataset contains approximately 504 training and 126 testing radiographs, with pneumoconiosis cases constituting about 63\% of each dataset.

\paragraph{Dataset pre-processing}
In the preprocessing phase, we first use the default window level and width (stored in the DICOM tags) to pre-process the original DICOM files, as in the approach described by \cite{wang2023real}. This step can adjust the contrast and brightness level of the chest radiographs, making the anatomical structure or lesion of interest easier to distinguish and observe. After the window adjustment, we use the pyplot.imsave function to convert the UInt16 DICOM format into the UInt8 PNG format. Next, we employ the CheXmask pipeline, introduced by \cite{gaggion2023chexmask}, for lung segmentation. Based on the lung segmentation results, we use the maximum external rectangle extraction technique to isolate the rectangular lung regions from the original chest radiographs. Finally, we resize these rectangular lung regions into a uniform size of 224×224 pixels for further analyses. Fig. \ref{fig_data_example} displays representative chest radiographs, showcasing both categories (e.g., pneumoconiosis and normal), the comparisons before and after the window adjustment operation, their corresponding lung segmentation results, and the extracted rectangular lung regions.

\paragraph{Evaluation metrics}
Following \cite{chen2023dynamyic} and \cite{qu2023generalized}, we adapt Accuracy (Acc.), Sensitivity (Sens.), Specificity (Spec.), and Area Under the Curve (AUC) for quantitative analysis to comprehensively evaluate the performance of pneumoconiosis diagnosis. To facilitate analysis and comparison, we calculate the average of these metrics to assess the overall performance.

\begin{figure*}[thbp]
\centering
\includegraphics[width=1\linewidth]{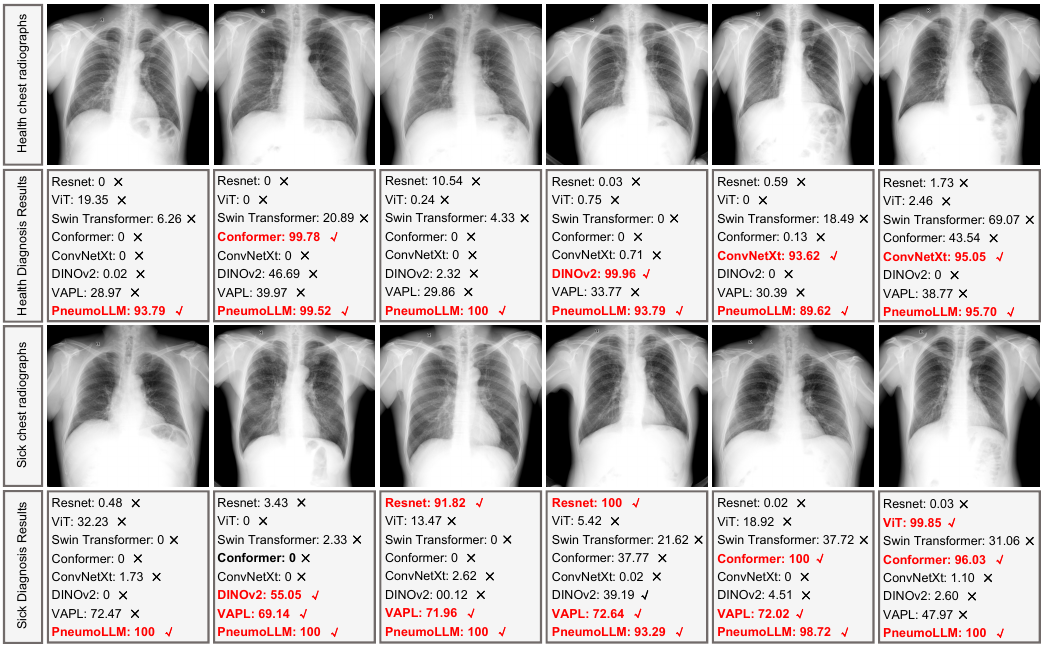}
\caption{Pneumoconiosis diagnosis results comparison with recent prestigious methods. The correct diagnosis results are highlighted in red.}
\label{fig_compare_fig}
\end{figure*}

\paragraph{Implementation details}
Consistent with \cite{luo2023cheap}, we employ the ViT-L/14 \citep{dosovitskiy2020image} of the pre-trained CLIP \citep{radford2021learning} as the vision encoder. We extract visual features as six \texttt{[CLS]} tokens from every fourth layer of ViT-L/14. The LLM utilize the LLaMA-7B \citep{touvron2023LLaMA} model. We set the visual neck dimension to 128 and the adapter dimension to 8. The adapter layer used in vision encoder and LLM is RepAdapter \citep{luo2023towards}. We employ AdamW \citep{loshchilov2017decoupled} as the optimizer, training the model for 100 epochs with a cosine decay learning rate schedule. The batch size, learning rate, warmup epochs, and weight decay are set to 16, 3$e^{-4}$, 2 and 0.02, respectively. Under this setting, due to the usage of frozen models and small trainable parameters, PneumoLLM fine-tuning is computationally efficient. For example, using a single NVIDIA GeForce RTX 4090 GPU with 24GB of memory, PneumoLLM requires less than 25 minutes for fine-tuning 100 epochs on a small-size training dataset of 504 chest radiographs.

\begin{figure*}[thbb]
\centering
\subfloat[ResNet]{\includegraphics[width=0.245\textwidth,trim=0.1in 0.1in 0.1in 0.1in, clip]{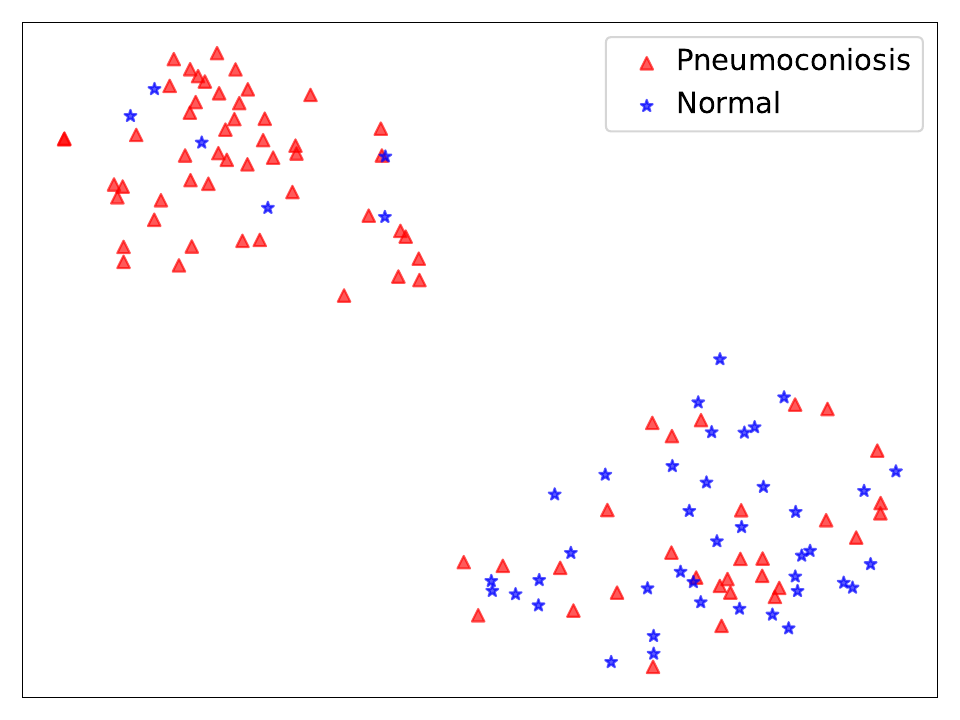}%
\label{fig_ResNet}}
\hfil
\subfloat[ViT]{\includegraphics[width=0.245\textwidth,trim=0.1in 0.1in 0.1in 0.1in, clip]{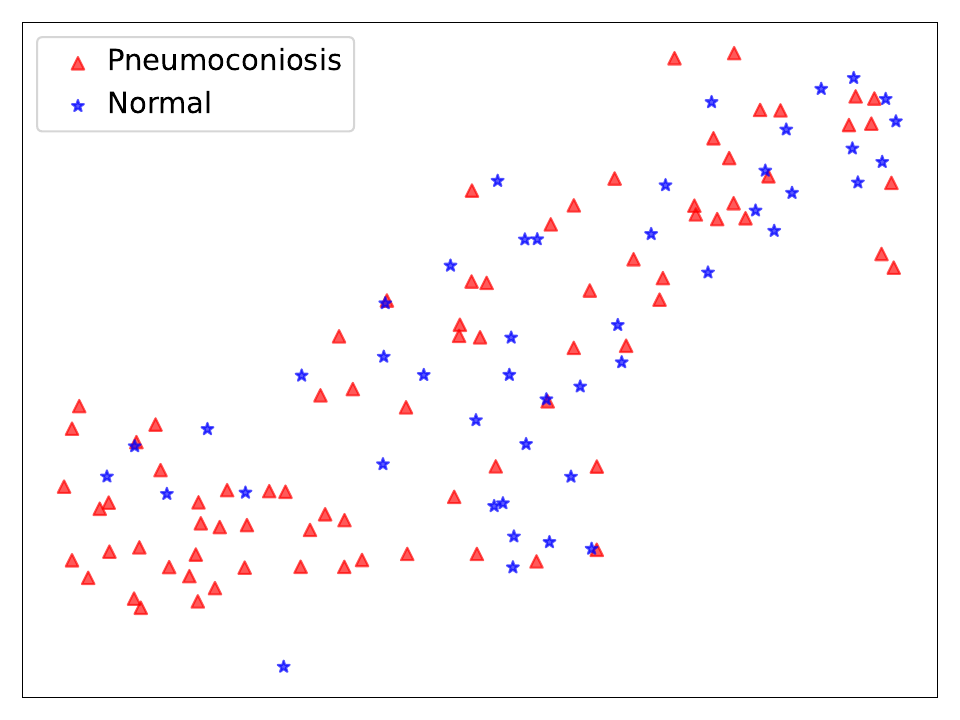}%
\label{fig_ViT}}
\hfil
\subfloat[Swin Transformer]{\includegraphics[width=0.245\textwidth,trim=0.1in 0.1in 0.1in 0.1in, clip]{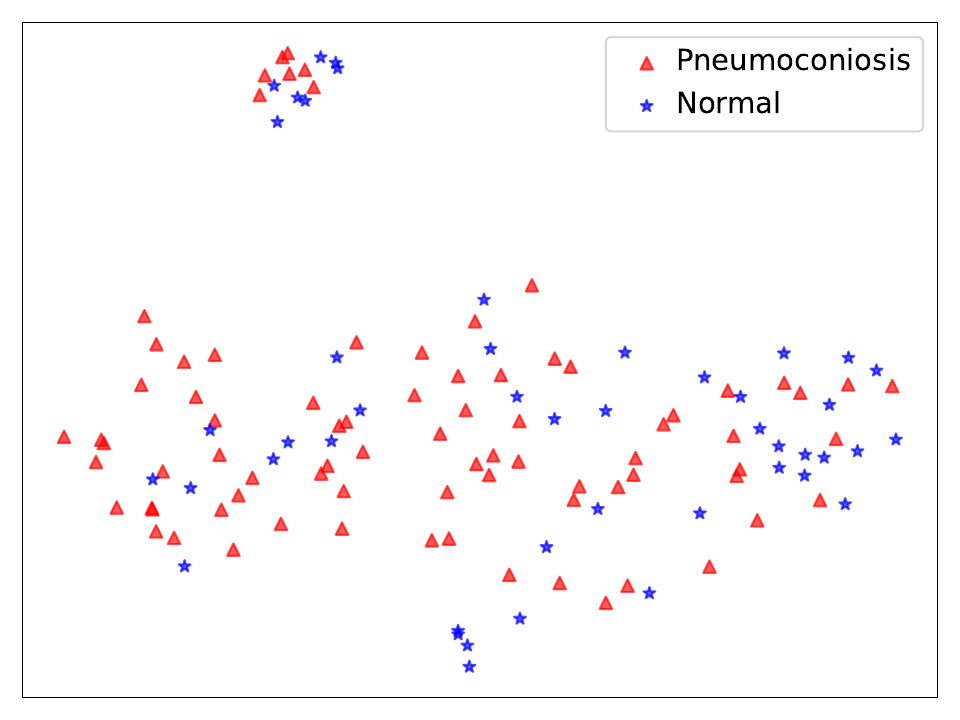}%
\label{fig_Swin}}
\hfil
\subfloat[Conformer]{\includegraphics[width=0.245\textwidth,trim=0.1in 0.1in 0.1in 0.1in, clip]{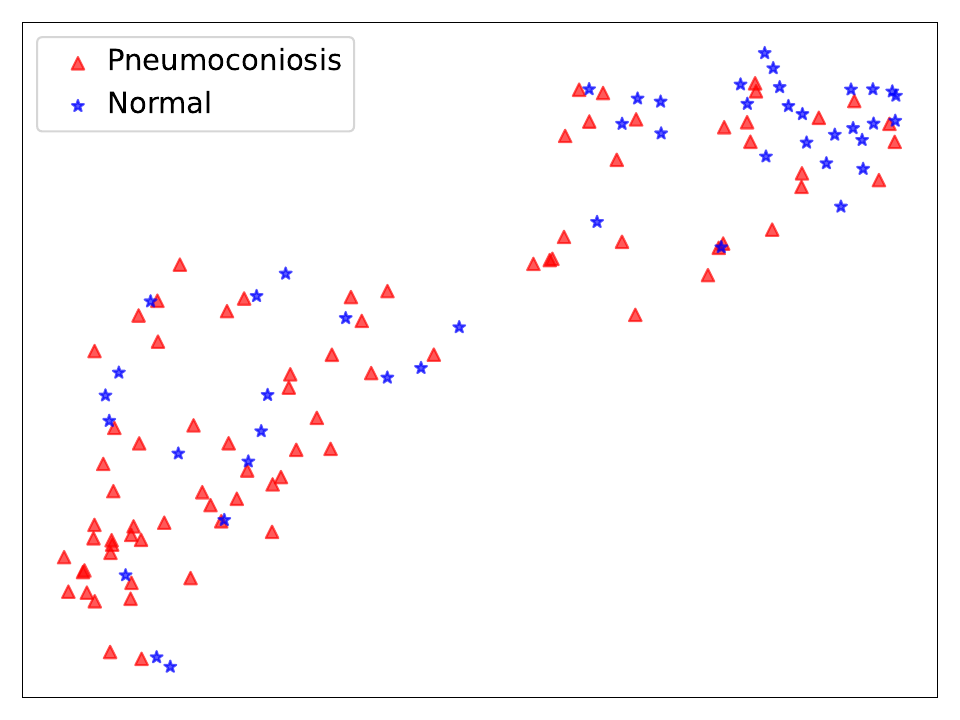}%
\label{fig_Conformer}}
\\
\subfloat[ConvNeXt]{\includegraphics[width=0.245\textwidth,trim=0.1in 0.1in 0.1in 0.1in, clip]{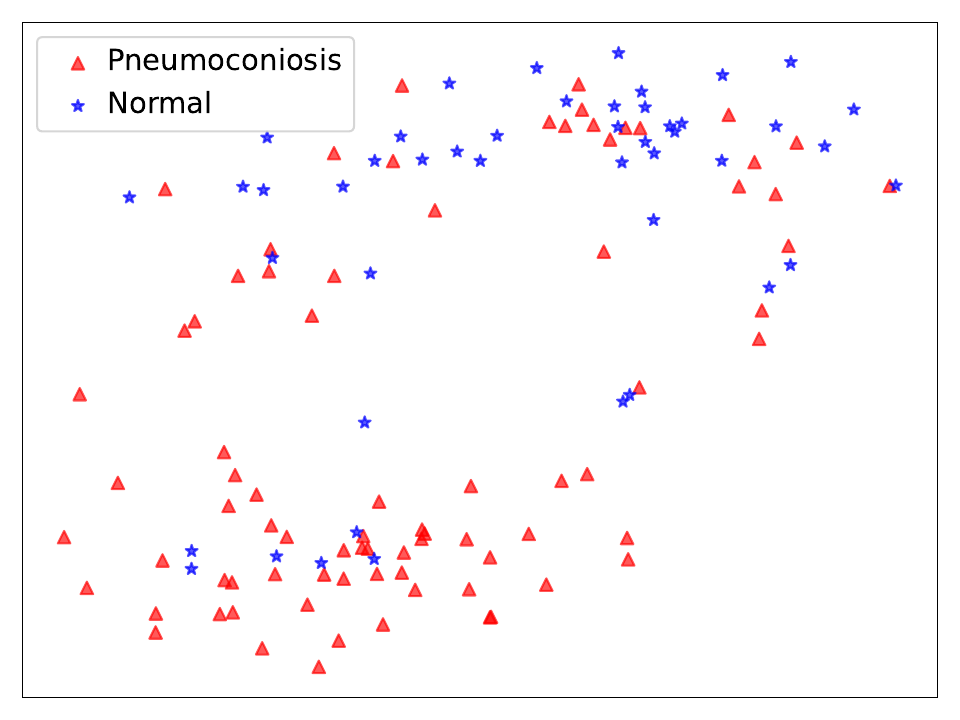}%
\label{fig_ConvNeXt}}
\hfil
\subfloat[DINOv2]{\includegraphics[width=0.245\textwidth,trim=0.1in 0.1in 0.1in 0.1in, clip]{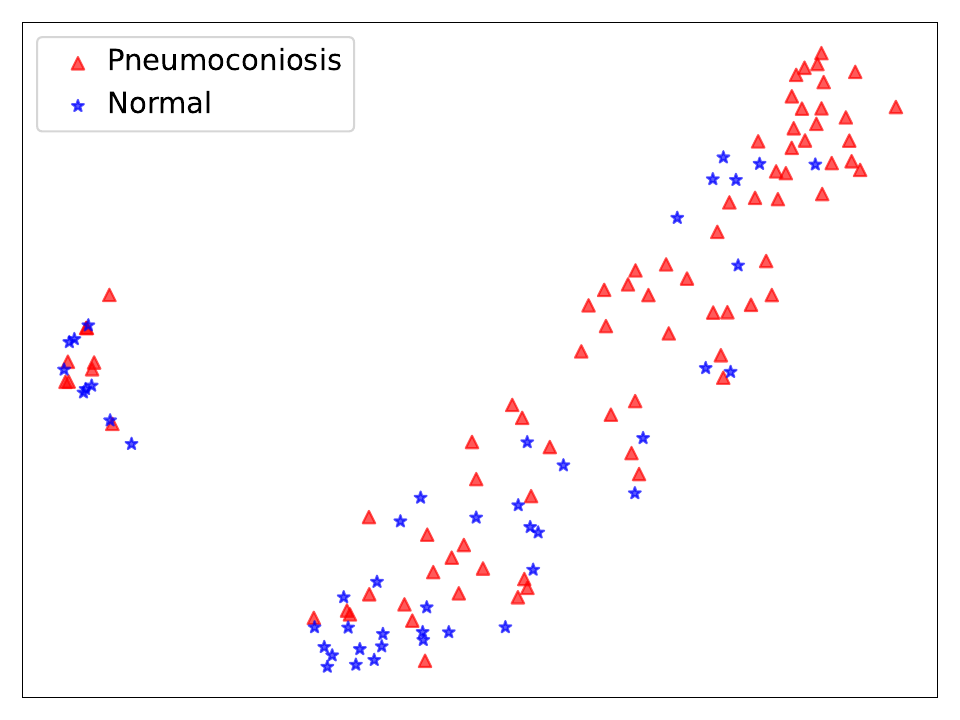}%
\label{fig_DINOv2}}
\hfil
\subfloat[VAPFormer]{\includegraphics[width=0.245\textwidth,trim=0.1in 0.1in 0.1in 0.1in, clip]{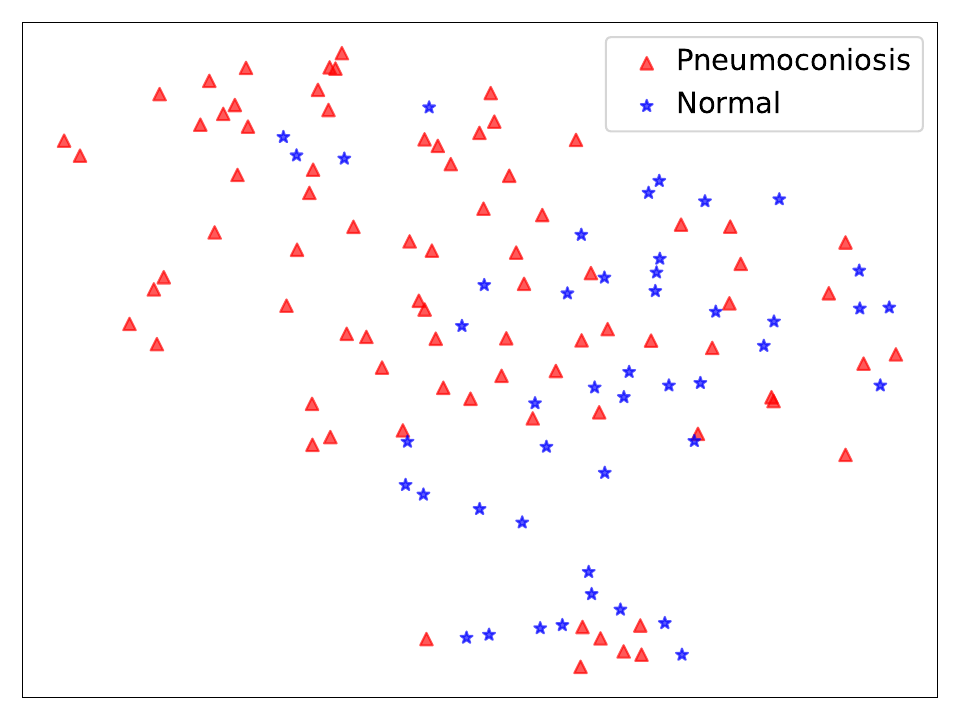}%
\label{fig_VAPFormer}}
\hfil
\subfloat[CoOp]{\includegraphics[width=0.245\textwidth,trim=0.1in 0.1in 0.1in 0.1in, clip]{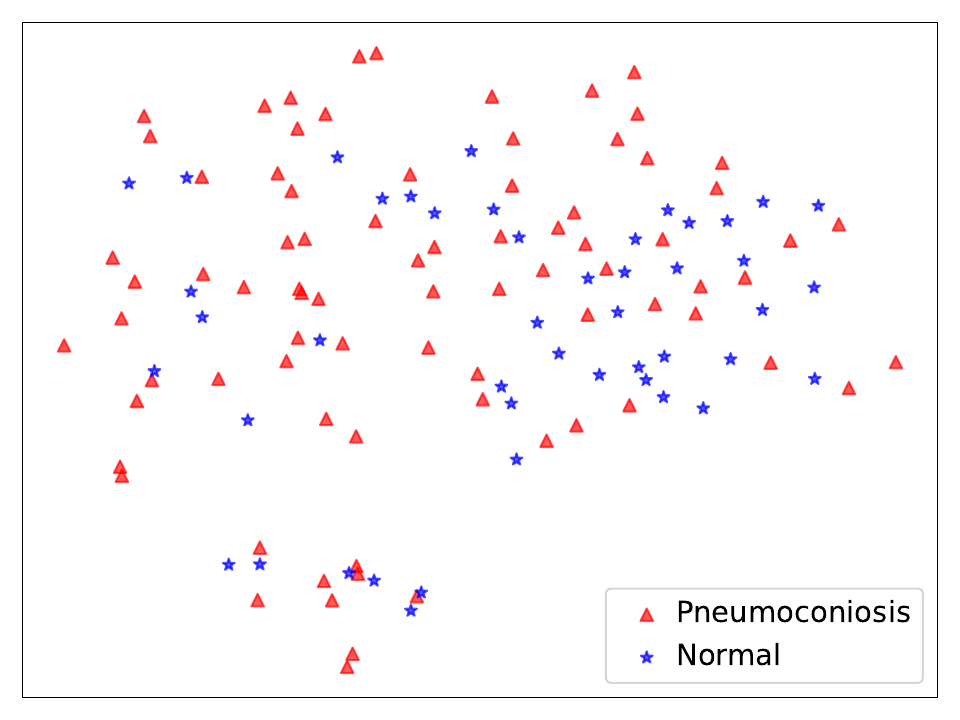}%
\label{fig_coop}}
\\
\hfil
\subfloat[CoCoOp]{\includegraphics[width=0.245\textwidth,trim=0.1in 0.1in 0.1in 0.1in, clip]{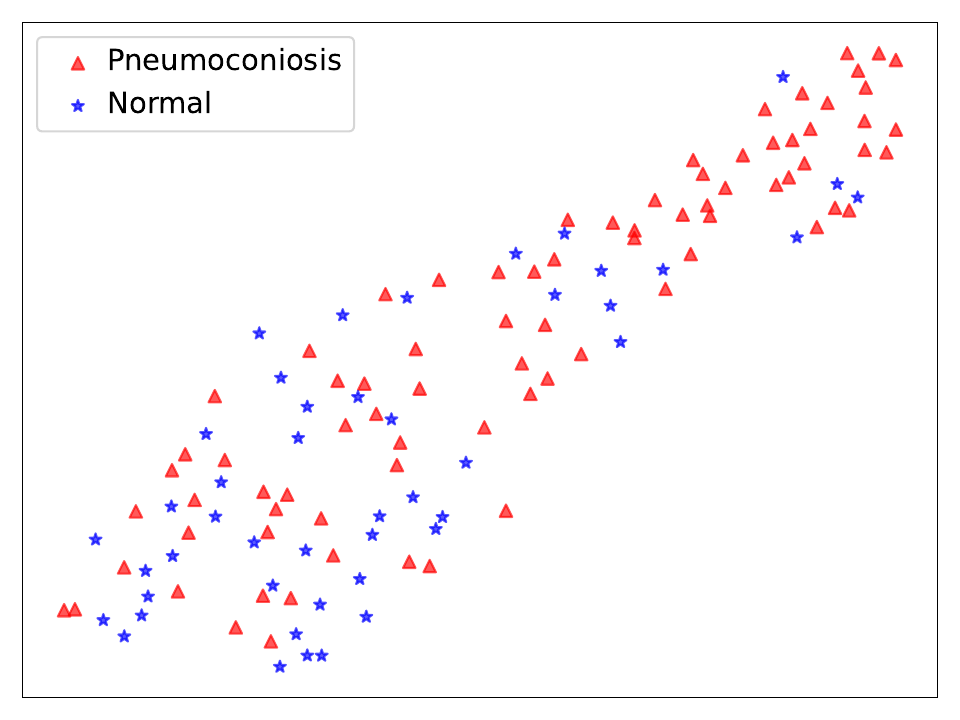}%
\label{fig_cocoop}}
\hfil
\subfloat[LaVIN]{\includegraphics[width=0.245\textwidth,trim=0.1in 0.1in 0.1in 0.1in, clip]{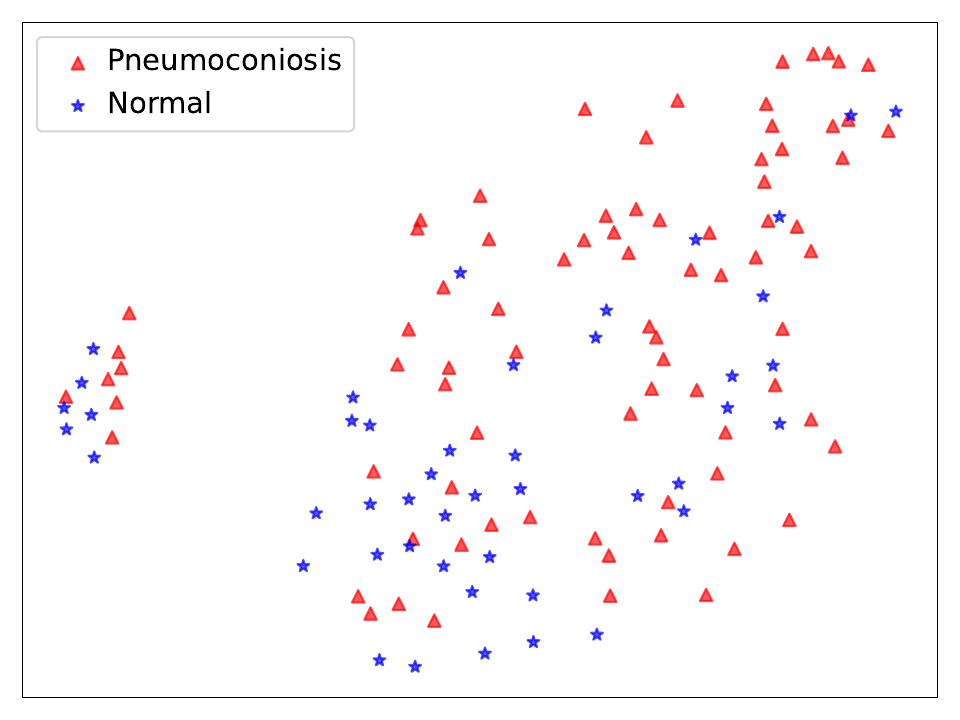}%
\label{fig_LAVIN}}
\hfil
\subfloat[BLIP2]{\includegraphics[width=0.245\textwidth,trim=0.1in 0.1in 0.1in 0.1in, clip]{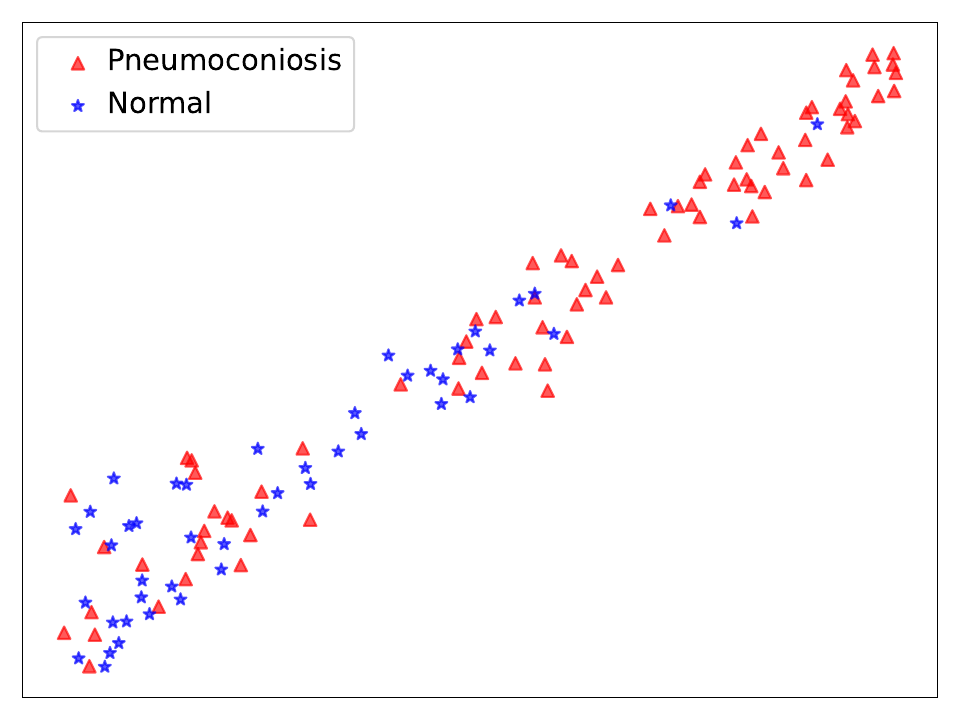}%
\label{fig_blip2}}
\hfil
\subfloat[Our PneumoLLM]{\includegraphics[width=0.245\textwidth,trim=0.1in 0.1in 0.1in 0.1in, clip]{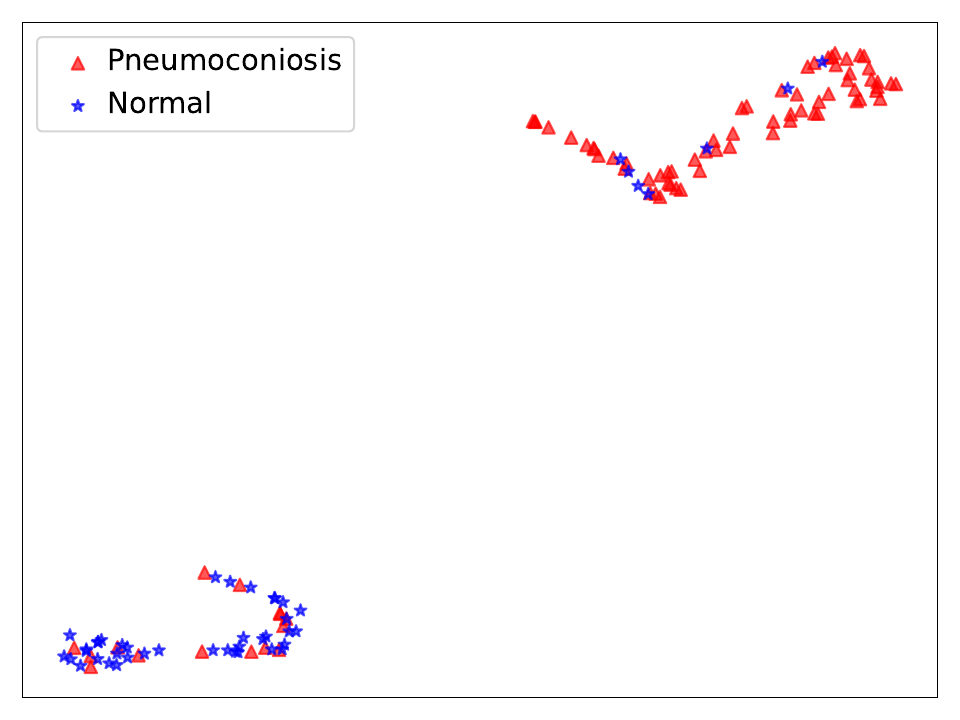}%
\label{fig_PneumoLLM}}
\caption{The t-SNE visualization of feature representation obtained by different networks in comparison experiment.}
\label{fig_tsne}
\end{figure*}
\subsection{Comparison experiments}
To evaluate our proposed PneumoLLM, we compare it against established natural image classification methods, including ResNet \citep{resnet}, ViT \citep{dosovitskiy2020image}, Swin Transformer \citep{liu2021swin}, Conformer \citep{conformer}, ConvNeXt \citep{convnext}, and DINOv2 \citep{dinov2}, as well as the advanced medical image diagnosis method VAPFormer \citep{kang2023visual}. All comparison models are implemented based on open-source configurations.

\paragraph{Pneumoconiosis diagnosis comparisons with recent prestigious methods} 
Tab. \ref{tabCmpExp} details the quantitative performance metrics of our PneumoLLM against the prestigious image classification methods which are pre-trained on ImageNet-1k \citep{deng2009imagenet} or LVD-142M \citep{dinov2}. Notably, among these prestigious image classification methods, models with inherent inductive biases like ResNet, Conformer, and ConvNeXt outperform transformer models like ViT and Swin Transformer in our limited Pneumoconiosis diagnosis dataset, while DINOv2 excels due to extensive pretraining data and prior information. Conversely, the medical image diagnosis method VAPFormer with prompts exhibits the least efficacy, which illustrates the big diagnostic challenges posed by data-scarce occupational disease. Our PneumoLLM, in constract, shows promising results by harnessing the power of LLM requiring relatively fewer trainable parameters. In addition, we can observe that all the methods in Tab. \ref{tabCmpExp} have significantly higher sensitivity and relatively lower specificity. The occurrence of this phenomenon is attributed to the unbalanced data distribution resulting from 401 out of the total 630 images in the dataset. 
Besides, we show some visual chest radiographs and performance comparisons by different algorithms in Fig. \ref{fig_compare_fig}. When compared with different algorithms to diagnose pneumoconiosis, PneumoLLM showed higher confidence in accurate predictions.

\paragraph{Pneumoconiosis diagnosis comparisons with recent foundational models}
Tab. \ref{tab_LLMcom} represents comparisons with recent existing foundational models, including Zero-Shot CLIP \citep{radford2021learning}, Linear Probe CLIP \citep{radford2021learning}, CoOp \citep{zhou2022learning} and CoCoOp \citep{zhou2022conditional}, representing vision-language contrastive learning methods (Fig. \ref{figMotivation}(a)), and BLIP2 \citep{li2023blip2} and LaVIN \citep{luo2023cheap}, representing vision-language alignment-based methods (Fig. \ref{figMotivation}(b)). Notably, for the vision-language contrastive learning methods, we set the text labels corresponding to normal and pneumoconiosis chest radiographs as `Normal chest radiograph' and `Pneumoconiosis chest radiograph'.  And for the vision-language alignment-based methods, we treat the pneumoconiosis diagnosis as a question-answering task with a coincident question input: ``Does the patient have pneumoconiosis?". We create `Yes' or `No' text answer labels to ensure consistency with the traditional visual question answering experiment setting.

As shown in Tab. \ref{tab_LLMcom}, directly using Zero-Shot CLIP on pneumoconiosis diagnosis tends to classify all chest radiographs as pneumoconiosis. This indicates that without pretraining, the vision features and language features extracted by CLIP are not discriminative due to the domain gap between natural images and medical chest radiographs. Even with a simple linear probe classifier on CLIP, the performance remains undesirable, as indicated by the very low AUC evaluation metric value. This further highlights the inability of frozen CLIP trained on natural images to learn a valid linear association with the target diagnosis label. In contrast, using simple context prompt learning strategies, like CoOp and CoCoOp, can partly alleviate this problem and greatly improve the performance of pneumoconiosis diagnosis. However, these strategies only affect the language feature representation in CLIP, and their performance improvement ability is limited, suggesting the need for further adaptation. 
Furthermore, the vision-language alignment-based methods, LaVIN and BLIP2, which use LLM to understand the image tokens extracted from the image encoder and text instruction tokens, perform better than CoOp and CoCoOp. This illustrates that LLM does have the ability to improve the generalization of vision representation. However, while the Qformer designed in BLIP2 performs better in mitigating this gap compared to the adapter usage in LaVIN, the relatively large number of learnable parameters and complex pretraining strategies makes it less efficient in harnessing LLM for the pneumoconiosis diagnostic task. There's still plenty of room to explore improving the ability of LLM to understand the image tokens extracted from the image encoder and perform better with small computing cost.  
In light of these challenges, our proposed PneumoLLM outperforms all of these methods. By directly eliminating the text branch and substituting the dialogue head with a classification head, along with the subtle contextual multi-token engine and information emitter module design, our method achieves a harmonious balance between preserving image representations and harnessing LLM's diagnostic intelligence. 
\begin{table*}[htbp]
  \centering
  \renewcommand\arraystretch{1}\setlength{\tabcolsep}{3.8mm}{%
  \caption{Analysis of LLaMA-7B foundational model in pneumoconiosis diagnosis.}
    \begin{tabular}{c|cc|cccc|c}
    \toprule
    Models & \multicolumn{1}{c}{\makecell{L-para.(M)}} & \multicolumn{1}{c|}{Mem. (G)}& 
    \multicolumn{1}{c}{Sens. (\%)} &
    \multicolumn{1}{c}{Spec. (\%)} &
    \multicolumn{1}{c}{Acc. (\%)} &
    \multicolumn{1}{c|}{AUC (\%)} &
    \multicolumn{1}{c}{AVG (\%)}\\
    \midrule
    \midrule
    {\makecell{PneumoLLM\\ w/o LLaMA}} & 0.52 & 4.15 &77.77 &62.00  & 72.06 & 75.81 &71.91 \\
    \midrule
    {\makecell{PneumoLLM}} & 2.70 & 15.48 &\textbf{80.54} &\textbf{67.66} & \textbf{75.87} & \textbf{78.98} &\textbf{75.76} (\textcolor{green!80!black}{+3.85}) \\
    \bottomrule
    \end{tabular}%
  \label{tab_abla_LLaMA}%
  }
\end{table*}%

\paragraph{Qualitative comparisons with t-SNE visualization}
To evaluate feature representation quality, we employ t-SNE \citep{van2008visualizing} method as qualitative comparisons to project high-level feature representations onto a 2D plane. We extracted the last-layer features and reshaped them into NCD-dimensional features, where N represents the number of samples, C denotes the channel dimension, and D represents the feature dimension. We then computed the mean along the channel dimension to obtain ND-dimensional features. This averaging operation collapses the channel information and generates a reduced representation for each sample. For CoOp and CoCoOp, since the decision criterion is based on the similarity comparison between visual features and text features of all categories, we concatenated all visual features and text features along the channel dimension and flattened them into ND-dimensional features. We then employed the t-SNE algorithm to project the ND-dimensional features onto a two-dimensional space for visualizing high-dimensional feature spaces. The final t-SNE visualization results are shown in Fig. \ref{fig_tsne}. 

Transformer-based architectures, like ViT, Swin Transformer, and LLM-based methods, like CoOp, do not perform optimally with limited datasets, exhibiting a less distinct separation between classes. In contrast, traditional convolutional networks like ResNet, along with hybrid models like Conformer and ConvNeXt, achieve better class discriminability. However, these models also present significant variance within class clusters. Notably, our proposed PneumoLLM model demonstrates a markedly superior clustering effect, with tightly grouped intra-class data points and stark demarcations between different classes. This compact and distinct representation suggests a more robust feature extraction capability. These visual insights are in concordance with our quantitative analyses, further corroborating the superior performance of PneumoLLM compared to conventional methodologies.

\subsection{Ablation experiments}
In our research, we perform a comprehensive ablation analysis of PneumoLLM, modifying one component at a time. Our PneumoLLM mainly comprises the utilization of LLM, contextual multi-token engine, and information emitter design. Initially, in Sec. \ref{sec:Model}, we analyze the model capacity, focusing on the essential role of the foundational LLaMA model, and the design of eliminating the textual processing branch by directly harnessing LLaMA to process visual features from the vision encoder. Subsequently, we evaluate the impact of various PneumoLLM components in Sec. \ref{sec:modules}, including the adapter, the contextual multi-token engine, and the information emitter module. 

\begin{table*}[htbp]
  \centering
  \renewcommand\arraystretch{1}\setlength{\tabcolsep}{4mm}{%
  \caption{Ablation study on eliminating the textual processing branch in LLM.}
    \begin{tabular}{c|cc|cccc|c}
    \toprule
    Settings & \multicolumn{1}{c}{\makecell{L-para.(M)}} & \multicolumn{1}{c|}{\makecell{Mem.(G)}}& \multicolumn{1}{c}{Sens. (\%)} & \multicolumn{1}{c}{Spec. (\%)} & \multicolumn{1}{c}{Acc. (\%)} & \multicolumn{1}{c|}{AUC (\%)} & \multicolumn{1}{c}{AVG (\%)} \\
    \midrule
    \midrule
    LaVIN  & 3.77 & 20.25 & \textbf{87.80}	& 47.52	& \textbf{73.33}	& 71.82	& 70.12\\
    \bottomrule
    {\makecell{Simplyfied\\PneumoLLM}}& 2.69 & 15.32 & 71.56 & \textbf{71.67} & 71.58 & \textbf{75.21} & \textbf{72.50} (\textcolor{green!80!black}{+2.38}) \\
    \hline
    \end{tabular}%
  \label{tab_ablation}%
  }
\end{table*}%

\subsubsection{Model capacity} 
\label{sec:Model}
\paragraph{Ablation study on LLaMA utilization}
In contrast to existing classification vision models, our PneumoLLM incorporates the LLaMA-7B model, harnessing its extensive knowledge and prior information to improve pneumoconiosis diagnosis. To validate this approach, we perform an ablation study examining the effects of LLaMA integration, detailed in Tab. \ref{tab_abla_LLaMA}. Our default PneumoLLM configuration integrates the ViT-L/14+LLaMA architecture with additional learnable adapter layers, contextual multi-token engine network, and classification network. Without LLaMA, PneumoLLM solely utilizes the ViT-L/14 architecture with adapter layers and classification network. As indicated in Table \ref{tab_abla_LLaMA}, the utilization of LLaMA demonstrates superior learning efficiency on small-size pneumoconiosis datasets, achieving improvements across all diagnosis metrics. However, this improvement comes at the cost of increased memory due to the LLM's inherent model size.

\paragraph{Ablation on eliminating the textual processing branch in LLM}
In contrast to LaVIN \citep{luo2023cheap}, which treats image classification as a question-answering task and requires both question text and image as input to the LLM, we argue that the question text inputs are redundant and unnecessary for the vision classification task. Instead, we propose that directly inputting the \texttt{[CLS]} tokens obtained from the pre-trained CLIP into LLaMA is sufficient to enhance pneumoconiosis diagnosis. To validate this assumption, we conduct an ablation study comparing our PneumoLLM model with image-only input (Fig. \ref{figMotivation}(c)) against LaVIN's dual image-question input setup (Fig. \ref{figMotivation}(b)), where the setup of LaVIN is consistent with the one described in Tab. \ref{tab_LLMcom}. For fairness, we build a simplified PneumoLLM by removing the contextual multi-token engine and information emitter module. The ablation results, presented in Tab. \ref{tab_ablation}, demonstrate that our simplified PneumoLLM outperforms LaVIN in terms of both efficiency and efficacy, supporting our hypothesis that the inclusion of question text input is unnecessary for pneumoconiosis diagnosis.

\begin{figure}[htbp]
\centering
\subfloat[Different vision encoder networks]{\includegraphics[height=1.32in]{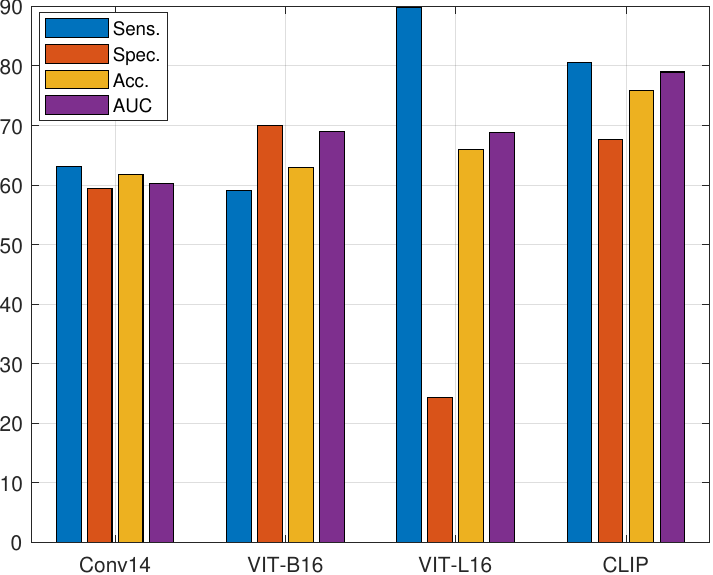}%
\label{fig_backbone_ab}}
\hfil
\subfloat[Number of generated diagnosis tokens]{\includegraphics[height=1.3in]{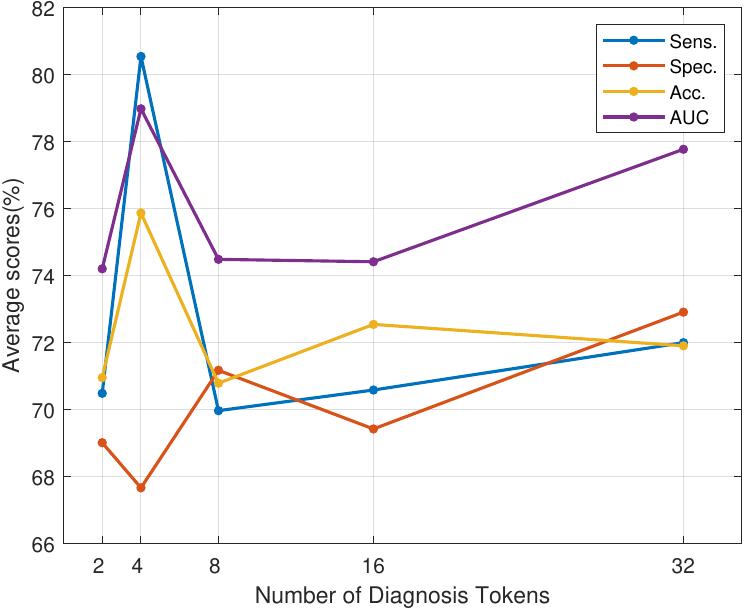}%
\label{fig_promote_number}}
\caption{Illustration on various vision encoder networks and the number of generated diagnosis tokens. Please zoom in for the best view.}
\label{fig_abla_exp}
\end{figure}

\begin{table*}[htbp]
  \centering
  \setlength\tabcolsep{3pt}   
  \renewcommand\arraystretch{1}\setlength{\tabcolsep}{1.6mm}{%
  \caption{Ablation study on various PneumoLLM components.}
    \begin{tabular}{cccccc|cccc|c}
    \toprule
    \multicolumn{1}{c}{Baseline} & 
    \multicolumn{1}{c}{Adapter} & 
    \multicolumn{1}{c}{CoOp} &
    \multicolumn{1}{c}{CoCoOp} &
    \multicolumn{1}{c}{\makecell{Contextual\\Multi-Token\\ Engine}} &
    \multicolumn{1}{c|}{\makecell{Information\\ Emitter}} &
    \multicolumn{1}{c}{Sens. (\%)} & \multicolumn{1}{c}{Spec. (\%)} & \multicolumn{1}{c}{Acc. (\%)} & \multicolumn{1}{c|}{AUC (\%)} & \multicolumn{1}{c}{AVG (\%)} \\
    \midrule
    \midrule
    \checkmark& & & & & & 78.07 & 62.77 & 72.54 & 74.52 & 71.98\\
    \checkmark&\checkmark & & & & & 71.56 & \textbf{71.67} & 71.58 & 75.21 & 72.50\\
    \checkmark&\checkmark&\checkmark& & & & 73.54 & 68.56 & 71.74 & 75.56 & 72.35\\
    \checkmark&\checkmark& &\checkmark & & & 72.57 & 70.31 & 71.75 & 75.80  & 72.61 \\
    \checkmark&\checkmark& & &\checkmark & & 78.29 & 66.82  & 74.13  & 76.42 & 73.92 \\
    \checkmark&\checkmark& & & \checkmark& \checkmark & \textbf{80.54} & 67.66 & \textbf{75.87} & \textbf{78.98} & \textbf{75.76}\\
    \bottomrule
    \end{tabular}%
    \label{tab_abla_learningstyle}%
    }
\end{table*}%

To further assess the impact of various vision encoders and the necessity of the usage of pre-trained CLIP network, we compare various vision encoders, including a standard $14 \times 14$ convolutional layer, ViT-B/16 and ViT-L/16 pre-trained on ImageNet-21k, and ViT-L/14 from pre-trained CLIP \citep{radford2021learning}. Results in Fig. \ref{fig_backbone_ab} indicate the superior performance of the ViT-L/14 from pre-trained CLIP. Notably, ViT-L/16 exhibited high sensitivity but low specificity, suggesting a tendency to over-diagnose. This ablation study highlights the significance of choosing appropriate pre-trained vision encoder for optimal integration with LLM.

\subsubsection{Ablation on various PneumoLLM components}
\label{sec:modules}

Due to the limited availability of pneumoconiosis patient data and patient privacy concerns, acquiring extensive pneumoconiosis datasets is challenging. Fine-tuning our entire PneumoLLM on the small-size pneumoconiosis dataset can make drastic changes in the LLM's parameter spaces, leading to increased training time and huge memory requirements. In this scenario, the usage of lightweight adapters, our proposed contextual multi-token engine and information emitter module prove crucial in adapting LLM efficiently to pneumoconiosis diagnosis. To validate this, we conduct comprehensive ablation studies in Tab. \ref{tab_abla_learningstyle}. In this table, we use the simplified PneumoLLM in Tab. \ref{tab_ablation} without adapters as our baseline for comparing various component configurations.

\paragraph{Ablation on adapter usage} 
We conduct an ablation study to measure the effect of the usage of adapters. The results, as presented in the first two rows of Tab. \ref{tab_abla_learningstyle}, indicate that the incorporation of adapters significantly enhances diagnostic performance in our small-size pneumoconiosis dataset.

\paragraph{Ablation on the contextual multi-token engine}
As discussed in Sec. \ref{sec:promote}, CoOp \citep{zhou2022learning} and CoCoOp \citep{zhou2022conditional} are two recent typical prompt engineering methods. To highlight the effectiveness of our proposed contextual multi-token engine design, we replace our contextual multi-token engine with the promote design in CoOp \citep{zhou2022learning} and CoCoOp \citep{zhou2022conditional} while keeping other components the same. The ablation results are in Tab. \ref{tab_abla_learningstyle}. While CoOp and CoCoOp settings show limited improvement, our multi-token engine yields a significant performance increase (+1.98\%).  CoOp's uniform prompts across samples hinder its adaptability, leading to a slight decrease in average performance. CoCoOp, though offering improved flexibility with vision-conditional prompts, faces limitations due to the mixed use of fixed and flexible prompts, yielding minimal gains. In contrast, our multi-token engine, designed to optimize the synergy between vision tokens and LLM, significantly enhances pneumoconiosis diagnosis.

\paragraph{Ablation on the information emitter module}
Another ablation study focuses on our proposed information emitter module. Results in Tab. \ref{tab_abla_learningstyle} demonstrate that combining the multi-token engine with the information emitter module achieves the best performance in pneumoconiosis diagnosis. This module successfully maintains the integrity of the original LLM’s information processing while emitting valuable information to context-diagnostic tokens, fostering enhanced diagnostic insights. 

Additionally, the optimal number of generated diagnosis tokens is determined based on performance analysis (Fig. \ref{fig_promote_number}), leading to the selection of four as the ideal number.

\section{Conclusion}
In this paper, we introduce PneumoLLM, a pioneering approach utilizing large language models for streamlined diagnostic processes in medical imaging. By discarding the text branch and transforming the dialogue head into a classification head, PneumoLLM simplifies the workflow for eliciting knowledge from LLMs. This innovation proves particular effectiveness when only classification labels are available for training, rather than extensive descriptive sentences. The streamlined process also significantly reduces the optimization space, facilitating learning with limited training data. Ablation studies further underscore the necessity and effectiveness of the proposed modules, especially in maintaining the integrity of source image details while advancing towards accurate diagnostic outcomes.

Still, our study has some limitations. First, our current work primarily explored how effectively leveraging LLMs can improve performance on the small pneumoconiosis diagnosis dataset. However, when dealing with multi-category and multi-label tasks, severe long-tailed classification problems may disrupt the model's learning process. Facing these more challenging complex scenarios, we need to consider more sophisticated improvements related to the realistic attributes of categories, such as incorporating causal multi-relationship graph designs when transferring source \texttt{[CLS]} tokens into diagnosis tokens in the contextual multi-token engine module. Additionally, the adapter layers and the final disease classification network may also require modifications to handle the increased complexity and diversity of diagnostic categories effectively. Looking ahead, we plan to expand PneumoLLM's application to more imaging modalities beyond chest radiography, e.g., CT and MRI scans, aiming to broaden its diagnostic capabilities across a spectrum of diseases. Second, while removing the text branch is successful in our current pneumoconiosis classification task which only focuses on the single vision modality, such an operation limits its application to more complex text-based multi-tasks in the medical domain, like multi-modality joint diagnosis and open-ended conversational generation \citep{li2024llava}. Therefore, in future research, we need to continue to find the better trade-off between computational complexity and the interactive fusion of features explored by different modalities. These future endeavors could enhance the capabilities of automated diagnostic systems, paving the way for more practical medical imaging analyses.

\section*{Acknowledgments}
This work was supported in part by grants from the Chinese Academy of Medical Sciences Innovation Fund for Medical Sciences, China  CIFMS2021-I2M-1-044, 2021-I2M-1-049, and the Non-profit Central Research Institute Fund of Chinese Academy of Medical Sciences, 2022-RC310-06, and the NSFC 62071382, 82272020, and the Basic and Applied Basic Research Foundation of Guangdong Province 2024A1515012388. Juntao Yang is a senior author to this work.

\section*{Declaration of generative AI and AI-assisted technologies in the writing process}
During the preparation of this work the authors used ChatGPT4.0 in order to improve readability and language of the work. After using this tool/service, the authors reviewed and edited the content as needed and take full responsibility for the content of the publication.

\bibliographystyle{model2-names.bst}\biboptions{authoryear}
\bibliography{refs}

\end{document}